\documentclass[12pt]{article}
 \usepackage[dvips]{graphicx}
  \usepackage{amsmath}
 \usepackage{amsxtra}
\begin{document}
\title {Generalized (s-Parameterized) Weyl Transformation}
\bigskip
\author{~ Alex Granik\thanks{Department of Physics,UOP,
Stockton,CA.95211;~E-mail:agranik@uop.edu}}
\date{}
\maketitle
\begin{abstract}
A general canonical transformation of mechanical operators of
position and momentum is considered. It is shown that it
automatically generates a parameter $s$ which leads to a
generalized ( or $s$-parameterized) Wigner function. This allows
one to derive a generalized ($s$-parameterized) Moyal brackets for
any dimensions. In the classical limit the $s$-parameterized
Wigner averages of the momentum and its square yield the
respective classical values. Interestingly enough, in the latter
case the classical Hamilton-Jacobi equation emerges as a
consequence of such a transition only if there is a non-zero
parameter $s$
\end{abstract}
\section{Introduction}
The  Moyal transformation in the context of the Weyl
transformation \cite{HW} (the former being a particular case of
the latter) was addressed by B.Leaf \cite{BL} who departing
directly from quantum mechanics derived   the Weyl-transform
$\it{A_w}({\bf{Q,K}})$ of the quantum operator $\it A({\bf{Q,K}})$
and investigated the properties of such a transform. Here
${\bf{Q}}$ and ${\bf{K}}$ denote the eigenvalues of the coordinate
${\bf{q}}$ and momentum ${\bf{k}}$ operators respectively. The
well-known Moyal formula for the phase-space distribution function
\cite{JM} readily followed from the derived expressions \cite
{BL}. However the work \cite{BL} did not yield a more general
$s$-parameterized transformation unifying different quantization
rules.\\

Usually this transformation is introduced "by hand" (cf.
\cite{CG},\cite{BJ}, \cite{DW}. A closer inspection of Leaf's
approach shows that it allows one to naturally arrive at the
generalized ( in a sense of $s$-parameterization)
Weyl-Wigner-Groenewold-Moyal transformation without a need for
$\it{apriori}$ introduction of the parameter $s$. This is
associated with the fact that a shift of the operators ${\bf{K}}$
and ${\bf{Q}}$ (${\bf{K,Q}} \rightarrow {\bf{K',Q'}}$) under the
only condition that the resulting transformation to be canonical
automatically generates an arbitrary parameter $s$ entering the
resulting transformation. In Ref. \cite{BL} this shift was chosen
in such a way as to satisfy the canonicity by a special choice of
the numerical coefficients entering the transformation ${\bf{K,Q}}
\rightarrow {\bf{K',Q'}}$ and ensuring the value of the parameter
$s=0.$

\section{Generalized Weyl Transformation}

Let us consider the Hilbert space of a quantum-mechanical system
having $n$ degrees of freedom. This space is spanned by the
eigenkets $|{\bf{Q}}>$ and $|{\bf{K}}>$ of the Cartesian
coordinate operator ${\bf{q}}(q_1,q_2,...,q_n)$ and the conjugate
momentum ${\bf{p}}= (\hbar/i)(\partial/\partial q)
=2\pi\hbar{\bf{k}}(k_1,k_2,...,k_n)$  [with the commutation
relations $ q_ik_j-k_iq_j=(i/2\pi)\delta_{ij};~i,j=1,2,...,n$ ]:
\begin{equation}\label{1}
{\bf{k}}|{\bf{K}}\rangle
={(2\pi\hbar)}^{-1}\frac{\hbar}{i}\frac{\partial}{\partial
q}|{\bf{K}}\rangle= {\bf{K}}|{\bf{K}}\rangle
\end{equation}
which means
\begin{equation}\label{2}
|{\bf{K}}\rangle=e^{2\pi i{\bf{q\bullet K}}};~~~~
|{\bf{Q}}\rangle=\delta(\bf{q-Q})
\end{equation}

The respective completeness relations are
\begin{equation}\label{3}
\ \int{d{\bf{Q}}|{\bf{Q}}\rangle \langle{\bf{Q}}|}=
{\bf{1}},~~~\int{d{\bf{K}}|{\bf{K}}\rangle \langle{\bf{K}}|}=
\bf{1}
\end{equation}

 Employing (1) and (2) we find that the scalar product
$<{\bf{Q}}|{\bf{K}}>$ is
\begin{equation}\label{4}
\langle{\bf{Q}}|{\bf{K}}\rangle=\int{e^{2\pi i{\bf{q\bullet
K}}}}\delta({\bf{q-Q}})d\bf{q}=e^{2\pi i{\bf{Q\bullet K}}}
\end{equation}\\

We represent an arbitrary quantum-mechanical operator $\it{A}$ in
the Hilbert space using the completeness relation (3):
\begin{equation}\label{5}
\it{A} \equiv
\int{...\int{d{\bf{Q'}}d{\bf{Q''}}d{\bf{K'}}d{\bf{K''}}|{\bf{Q''}}
\rangle\langle{\bf{K''}}|{\bf{K''}}\rangle\langle{\bf{K''}}|\it{A}|{\bf{K'}}\rangle
\langle{\bf{K'}}|{\bf{Q'}}\rangle \langle{\bf{Q'}}|}}
\end{equation}
Now we perform a linear transformation from the variables
$\bf{Q',Q'',K',K''}$ to new variables $\bf{K,Q,u,v}$ according to
the following
\begin{eqnarray}\label{6,7}
Q_i''=Q_i+\alpha_iv_i,~~~~Q_i'=Q_i+\beta_iv_i\\
K_i''=K_i+\gamma_iu_i,~~~~K_i'=K_i+\delta_iu_i; ~~~~i=1,2..,.n
\end{eqnarray}
where $\alpha_i,\beta_i,\gamma_i$ and $\delta_i$ are some
constants to be determined from an additional condition. If we
require this transformation to be $canonical$ then its Jacobian
must be $1$ yielding the following relations:
\begin{equation}\label{8}
\prod_{i=1}^{i=n}(\alpha_i - \beta_i)(\gamma_i-\delta_i) =1
\end{equation}
Note that in \cite{BL} from the very beginning the coefficients
are chosen in such a way as to identically satisfy the canonicity
condition: $ \alpha_i=\gamma_i =1/2; ~\beta_i=\delta_i=-1/2$.\\

We rewrite identity (5) taking into account the transformation of
variables (6),(7):

\begin{eqnarray}\label{9}
\it A \equiv \int\dots \int d {\bf{Q}} d{\bf{K}}
d{\bf{u}}d{\bf{v}} \mid{\bf{Q}}+ \alpha\bf{v}\rangle\langle
\bf{Q}+ \alpha \bf{v} \mid
\bf{K}+ \gamma \bf{u}\rangle \nonumber \\
\times\langle \bf{K}+\gamma\bf{u}\mid \it A \mid
\bf{K}+\delta\bf{u} \rangle \langle \bf{K}+ \delta \bf{u}|
 \bf{Q}+ \beta \bf{v}\rangle \langle \bf{Q}+ \beta \bf{v}\mid
\end{eqnarray}

where according to (4)
\begin{eqnarray*}
\lefteqn{\langle \bf{Q}+\alpha\bf{v}\mid
\bf{K}+\delta\bf{u} \mid \bf{Q}+\beta\bf{v}\rangle =} \\
& exp\{2\pi i[\sum_{i=1}^{n}(\gamma_i-\delta_i)u_iQ_i+
(\alpha_i-\beta_i)v_iK_i+u_iv_i
(\alpha_i\gamma_i-\beta_i\delta_i)]\}
\end{eqnarray*}
 Inserting this expression into Eq.(9) we get
 \begin{eqnarray}\label{10}
\it A \equiv \int \dots \int d{\bf{Q}} d{\bf{K}} d{\bf{u}}
 d{\bf{v}}  \langle \bf{K}+\gamma\bf{u} \mid \it A \mid
\bf{K}+\delta\bf{u} \rangle\mid Q+\alpha \bf{v}\rangle \langle
\bf{Q} +\beta \bf{v} \mid \nonumber \\
exp\{2\pi i[\sum_{i=1}^{n}(\gamma_i-\delta_i)u_iQ_i+
(\alpha_i-\beta_i)v_iK_i+u_iv_i
(\alpha_i\gamma_i-\beta_i\delta_i)]\}
\end{eqnarray}

By a simple change of variables we can incorporate
$u_iv_i(\alpha_i\gamma_i-\beta_i\delta_i)$ into variables $Q_i$.
To this end we represent the power of the exponent in (10)as
follows
\begin{eqnarray}\label{11}
(\gamma_i-\delta_i)u_iQ_i+ (\alpha_i-\beta_i)v_iK_i+u_iv_i
(\alpha_i\gamma_i-\beta_i\delta_i)= \nonumber \\
(\gamma_i-\delta_i)u_i[Q_i+ v_i
\frac{\alpha_i\gamma_i-\beta_i\delta_i}{\gamma_i-\delta_i}]+(\alpha_i-\beta_i)v_iK_i
\end{eqnarray}
Introducing a new variable $$Q^o_i =Q _i+ v_i
\frac{\alpha_i\gamma_i-\beta_i\delta_i}{\gamma_i-\delta_i},$$
dropping the superscript $o$, and using the fact that
$$exp\{-2\pi{\bf{k}}\bullet{\bf{a}}\}\mid {\bf{Q}}>=\mid {\bf{Q
+a}}>$$ we obtain the following representation of the operator
$\it A$:
\begin{equation}\label{12}
{\it{A}}=\int \dots \int d{\bf{Q}}d{\bf{K}}{\it
{A}}_w(\gamma,\delta,{\bf{Q}},{\bf{K}})
{\it{F}}(\alpha,\beta,{\bf{Q}},{\bf{K}})
\end{equation}
where
\begin{equation}\label{13}
\it {A}_w(\gamma,\delta,{\bf{Q}},{\bf{K}})=\int d{\bf{u}}e^{2\pi
i\sum_{k=1}^{n}(\gamma_k-\delta_k)u_kQ_k} \langle {\bf{K}}+\gamma
{\bf{u}}\mid\it{A}\mid {\bf{K}}+\delta {\bf{u}}\rangle,
\end{equation}

\begin{equation}\label{14}
\it{F}(\alpha,\beta,r,{\bf{Q}},{\bf{K}})=\int d{\bf{v}}e^{-2\pi
i(\alpha-\beta){\bf{v(k-K)}}}\mid{\bf{Q+rv}}\rangle\langle{\bf{Q+rv}}\mid
\end{equation}
and $${\bf{r}}=r_i=(r_1,r_2,...,r_n)=
\frac{\gamma_i}{\gamma_i-\delta_i}(\beta_i-\alpha_i)$$\\

The Weyl transformation is characterized by the properties of the
operator $\it F(\alpha,\beta,r,{\bf{Q}},{\bf{K}}).$ To
investigate it further we  use the following representation of the
projection operator $\mid {\bf{Q +rv}}\rangle\langle{\bf{Q +r
v}}\mid$:
\begin{eqnarray}\label{15}
\mid {\bf{Q +rv}}\rangle\langle{\bf{Q +rv}}\mid=
\delta({\bf{q-Q-r v}})=\nonumber \\
\int d{\bf{w}}e^{2\pi i\sum
w_j(q_j-Q_j)}e^{-2\pi i\sum r_jw_jv_j}
\end{eqnarray}

As a result we get from (\ref{14})
\begin{equation}\label{16}
{\it{F}}(\alpha,\beta,r,{\bf{Q}},{\bf{K}})=\int d{\bf{w}}e^{2\pi
i\sum w_j(q_j-Q_j)}e^{-2\pi i\sum r_jw_jv_j}\int
d{\bf{v}}e^{-2\pi i \sum(\alpha_j-\beta_j)v_j(k_j-K_j)}
\end{equation}
Because both $\bf{q}$ and $\bf{k}$ are hermitian the last
expression demonstrates that
${\it{F}}(\alpha,\beta,r,{\bf{Q}},{\bf{K}})$ is also
Hermitian.
\\

Since $$(\frac{\partial}{\partial Q_j})^m\int d{\bf{w}}e^{2\pi
i\sum w_j(q_j-Q_j)}= (-2\pi i)^m\int d{\bf{w}}(w_j)^m e^{2\pi
i\sum w_j(q_j-Q_j)}$$ and
\begin{eqnarray*}
(\frac{\partial}{\partial K_j})^m\int d{\bf{v}}e^{-2\pi
i\sum(\alpha_j-\beta_j)v_j(k_j-K_j)}=\\
(2\pi i)^m(\alpha_j-\beta_j)^m\int d{\bf{v}}(v_j)^m e^{-2\pi i
\sum(\alpha_j-\beta_j)v_j(k_j-K_j)}
\end{eqnarray*}
Eq.(16) yields
\begin{eqnarray}\label{17}
{\it{F}}(\alpha,\beta,r,{\bf{Q}},{\bf{K}})= \nonumber \\
e^{\frac{1}{2\pi i}\sum
\frac{r_j}{\alpha_j-\beta_j}\frac{\partial}{\partial
Q_j}\frac{\partial}{\partial K_j}}\int d{\bf{v}}e^{-2\pi i
\sum(\alpha_j-\beta_j)v_j(k_j-K_j)}\int d{\bf{w}}e^{2\pi i\sum
w_j(q_j-Q_j)}
\end{eqnarray}

By introducing new variables
$k_j^o=k_j(\alpha_j-\beta_j),K_j^o=K_j(\alpha_j-\beta_j)$
 the parameters $(\alpha_j-\beta_j)$ are "absorbed" by these variables,
which means that without any loss of generality we can set
$$(\alpha_j-\beta_j)=1.$$ Therefore the operator
$\it{F}(\alpha,\beta,r,{\bf{Q}},{\bf{K}})$ becomes
\begin{equation}\label{18}
\it{F}({\bf{r,Q,K}})=e^{\frac{1}{2\pi i}\sum
r_j\frac{\partial}{\partial Q_j}\frac{\partial}{\partial
K_j}}\int d{\bf{v}}e^{-2\pi i \sum v_j(k_j-K_j)}\int
d{\bf{w}}e^{2\pi i\sum w_j(q_j-Q_j)}
\end{equation}
Since $$\int d{\bf{w}}e^{2\pi i\sum
w_j(q_j-Q_j)}=\delta({\bf{q-Q}}); ~~~\int d{\bf{v}}e^{-2\pi i\sum
v_j(k_j-K_j)}=\delta({\bf{k-K}})$$ we obtain from Eq.(18) another
representation of $\it F$
\begin{equation}\label{19}
\it{F}({\bf{r,Q,K}})=e^{\frac{1}{2\pi i}\sum
r_j\frac{\partial^2}{\partial Q_j \partial
K_j}}\delta({\bf{q-Q}})\delta{\bf{k-K}})
\end{equation}

The last equation allows us to find an explicit expression for the
Weyl-transform of the quantum operator
${\it{A_w}}({\bf{\gamma,Q,K}})$ which we represent as follows:
\begin{equation}\label{20}
\it{A_w}({\bf{r,Q',K'}})=\int
d{\bf{Q}}d{\bf{K}}\it{A_w}({\bf{r,Q,K}})\delta({\bf{Q-Q'}})\delta({\bf{K-K'}})
\end{equation}
Using (19) we calculate $<{\bf{Q'}}\mid {\it
{F}}(\bf{r,Q,K})\mid{\bf K'}>$:
\begin{equation}\label{21}
\langle{\bf{Q'}}\mid {\it {F}}(\bf{r,Q,K})\mid{\bf
K'}\rangle=\langle {\bf{Q'}}\mid{\bf{K'}}\rangle e^{\frac{1}{2\pi
i}\sum r_j \frac{\partial^2}{\partial Q_j\partial
K_j}}\delta({\bf{Q'-Q}})\delta({\bf{K'-K}}) \end{equation}

where we use the following identity
$$\mid{\bf{Q}}\rangle\langle{\bf{Q}}\mid{\bf{Q'}}\rangle\equiv
\delta({\bf{Q'-Q}})\mid{\bf{Q'}}\rangle \equiv \delta({\bf{q}}-{\bf{Q}})
\mid {\bf{Q'}}\rangle$$\\

With the help of another identity
$\langle{\bf{Q'}}\mid{\bf{K'}}\rangle\langle{\bf{K'}}\mid{\bf{Q'}}\rangle
\equiv 1$ we get from (21)
\begin{equation}\label{22}
\delta({\bf{Q'}}-{\bf{Q}})\delta({\bf{K'}}-{\bf{K}})=
e^{-\frac{i}{2\pi}\sum r_j \frac{\partial^2}{\partial Q'_j\partial
K'_j}}\langle{\bf{Q'}}\mid \it{F}({\bf{r,Q,K}})\mid
{\bf{K'}}\rangle\langle{\bf{K'}}\mid{\bf{Q'}}\rangle
\end{equation}
Substitution of (22) into (20) yields
\begin{eqnarray}
\label{23}
\it{A}_w({\bf{r,Q',K'}})=\nonumber \\
\int
d{\bf{Q}}d{\bf{K}}{\it{A}}_w({\bf{r,Q,K}})e^{-\frac{i}{2\pi}\sum
r_j \frac{\partial^2}{\partial Q'_j\partial
K'_j}}\langle{\bf{Q'}}\it{F}({\bf{r,Q,K}})\mid
{\bf{K'}}\rangle\langle{\bf{K'}}\mid{\bf{Q'}}\rangle
\end{eqnarray}\\

On the other hand, from (12) follows that
$$\langle{\bf{Q'}}|\it{A}|{\bf{K'}}\rangle\langle{\bf{K'}}|{\bf{Q'}}\rangle=\int
d{\bf{Q}}d{\bf{K}}\it{A}_w({\bf{r,Q,K}})\langle{\bf{Q'}}|
F({\bf{r,Q,K}})|
{\bf{K'}}\rangle\langle{\bf{K'}}|{\bf{Q'}}\rangle.$$ Therefore
\begin{eqnarray}\label{24}
e^{-\frac{i}{2\pi}\sum r_j\frac{\partial^2}{\partial Q'_j\partial
K'_j}}\langle{\bf{Q'}}|\it{A}|{\bf{K'}}
\rangle\langle{\bf{K'}}|{\bf{Q'}}\rangle= \nonumber \\
\int d{\bf{Q}}d{\bf{K}}\it{A}_w(\bf{r,Q,K}) e^{-\frac{i}{2\pi}\sum
r_j\frac{\partial^2}{\partial Q'_j\partial
K'_j}}\langle{\bf{Q'}}|\it{F}({\bf{r,Q,K}})\mid
{\bf{K'}}\rangle\langle{\bf{K'}}|{\bf{Q'}}\rangle
\end{eqnarray}
Combining (23) and (24) we get the following expression for
$\it{A}_w({\bf{r,Q,K}})$:
\begin{equation}\label{25}
\it{A}_w({\bf{r,Q,K}})=e^{-\frac{i}{2\pi}\sum
r_j\frac{\partial^2}{\partial Q'_j\partial
K'_j}}\langle{\bf{Q'}}|\it{A}|{\bf{K'}}\rangle\langle{\bf{K'}}|{\bf{Q'}}\rangle
\end{equation}\\

If we  replace $K_j\rightarrow P_j/2\pi\hbar$ and take into
account that for a system with $N$ degrees of freedom
$\mid{\bf{K}}\rangle  = (2\pi\hbar)^{-N/2}\mid{\bf{P}}\rangle$
then (\ref{25}) takes the following form:
\begin{eqnarray}\label{26}
\it{A}_w({\bf{r,Q',P'}})=(2\pi\hbar)^Ne^{-i \hbar\sum
r_j\frac{\partial^2}{\partial P'_j\partial
K'_j}}\langle{\bf{Q'}}\mid\it{A}\mid{\bf{P'}}\rangle\langle{\bf{P'}}\mid{\bf{Q'}}\rangle=\nonumber
\\(2\pi\hbar)^Ne^{i \hbar\sum
r_j\frac{\partial^2}{\partial Q'_j\partial
P'_j}}\langle{\bf{Q'}}\mid{\bf{P'}}\rangle\langle{\bf{P'}}\mid\it{A}\mid{\bf{Q'}}\rangle
\end{eqnarray}
To express Eq.(26) in the Schroedinger representation we consider
an orthonormal set of eigenkets $\mid\psi_m\rangle$ and expand the
operator $\it{A}$ in terms of these eigenkets( eigenbras)
\begin{equation}\label{27}
\it{A}=\sum_{mn} w_m\mid\psi_m\rangle w^{*}_n\langle\psi_n\mid
\end{equation}
where $w_m$ are the respective coefficients of the expansion.
Upon substitution of (27) into (26) we obtain the following
expression
\begin{equation}
\label{28} \it{A}_w({\bf{r,Q,P}})=e^{i \hbar\sum
r_j\frac{\partial^2}{\partial Q_j\partial
P_j}}[\psi^*({\bf{Q}})\psi({\bf{P}})e^{i{\bf{PQ/\hbar}}}]
\end{equation}
which is the generalization of the Moyal formula (expression
(3.10) in \cite{JM}) which follows from \eqref{28} for the
particular value of the
parameter $r_j=-1/2,~ j=1,2,...,N.$\\

As a next step, we find the commutator $[A,B]$ of operators $A$
and $B$ from the expression (16) for $F({\bf{r,Q,K}})$ where we
use $\alpha_j-\beta_j =1$. With  the help of the Baker-Hausdorf
identity for two operators whose commutator is a constant
$$e^{\it{A+B}}e^{[\it{A,B}]}=e^Ae^{\it{B}}$$ we rewrite (16)
$$F({\bf{r,Q,K}})=\int d{\bf{v}}d{\bf{w}}e^{-2\pi i\sum
r_jw_jv_j}e^{2 \pi i\sum w_j(q_j-Q_j)}e^{-2\pi i\sum
v_j(k_j-K_j)}=$$ $$ \int d{\bf{v}}d{\bf{w}}e^{-\pi
i\sum(1+2r_j)w_jv_j}e^{-2\pi i\sum(w_jQ_j-v_jK_j)}e^{2\pi
i\sum(w_jq_j-v_jk-j)}$$

 This expression yields
\begin{eqnarray}\label{29}
F({\bf{r,Q,K}})\delta({\bf{Q'-Q}})\delta({\bf{K'-K}}) =\nonumber \\
\int d{\bf{v}}d{\bf{w}}d{\bf{v'}}d{\bf{w'}}e^{2\pi i\sum
w_jq_j-v_jk_j}e^{-2\pi
i\sum[Q_j(w'_j+w/2)+K_j(v'_j-v/2)]}\times\nonumber \\
e^{2\pi i\sum[Q'_j(w'-w/2)+K'(v'+v/2)]}
\end{eqnarray}
Introducing new variables $$v''=v'+v/2;~~ v'''=-(v-v/2);$$
$$w''=-(w'-w/2);~~w'''=w'+w/2$$ we obtain after some (rather
lengthy)
algebra$$F({\bf{r,Q,K}})\delta({\bf{Q'-Q}})\delta({\bf{K'-K}})=
e^{-(i/4\pi) \sum[\frac{\partial}{\partial
Q_j}\frac{\partial}{\partial K'_j}-\frac{\partial}{\partial
K_j}\frac{\partial}{\partial Q'_j}]}\times$$ $$
e^{-(i/4\pi)\sum(1+2 r_j) [\frac{\partial}{\partial
Q_j}\frac{\partial}{\partial K'_j}+\frac{\partial}{\partial
K_j}\frac{\partial}{\partial
Q'_j}]}F({\bf{r,Q,K}})F({\bf{r,Q',K'}})$$

 Using this expression we readily obtain that
$$F({\bf{r,Q,K}})F({\bf{r,Q',K'}})=e^{(i/4\pi)
\sum[\frac{\partial}{\partial Q_j}\frac{\partial}{\partial
K'_j}-\frac{\partial}{\partial K_j}\frac{\partial}{\partial
Q'_j}]}\times$$ $$ e^{(i/4\pi)\sum(1+2 r_j)
[\frac{\partial}{\partial Q_j}\frac{\partial}{\partial
K'_j}+\frac{\partial}{\partial K_j}\frac{\partial}{\partial
Q'_j}]}F({\bf{r,Q,K}})\delta({\bf{Q'-Q}})\delta({\bf{K'-K}})$$

Now we can write the product of two $AB$ operators as follows
\begin{eqnarray}\label{30}
\it{A B}=\int
d{\bf{Q}}d{\bf{K}}d{\bf{Q'}}d{\bf{K'}}\it{A}_w({\bf{r,Q,K}})\it{B}_w({\bf{r,Q',K'}})
F({\bf{r,Q,K}})F({\bf{r,Q',K'}})=\nonumber \\
\int
d{\bf{Q}}d{\bf{K}}d{\bf{Q'}}d{\bf{K'}}\it{A}_w({\bf{r,Q,K}})\it{B}_w({\bf{r,Q',K'}})e^{(i/4\pi)
\sum[\frac{\partial}{\partial Q_j}\frac{\partial}{\partial
K'_j}-\frac{\partial}{\partial K_j}\frac{\partial}{\partial
Q'_j}]}\times\nonumber \\ e^{(i/4\pi)\sum(1+2 r_j)
[\frac{\partial}{\partial Q_j}\frac{\partial}{\partial
K'_j}+\frac{\partial}{\partial K_j}\frac{\partial}{\partial
Q'_j}]}F({\bf{r,Q,K}})\delta({\bf{Q'-Q}})\delta({\bf{K'-K}})=\nonumber
\\ \int
d{\bf{Q}}d{\bf{K}}d{\bf{Q'}}d{\bf{K'}}F({\bf{r,Q,K}})\delta({\bf{Q'-Q}})
\delta({\bf{K'-K}})e^{(i/4\pi) \sum[\frac{\partial}{\partial
Q_j}\frac{\partial}{\partial K'_j}-\frac{\partial}{\partial
K_j}\frac{\partial}{\partial Q'_j}]}\times\nonumber \\
e^{(i/4\pi)\sum(1+2 r_j) [\frac{\partial}{\partial
Q_j}\frac{\partial}{\partial K'_j}+\frac{\partial}{\partial
K_j}\frac{\partial}{\partial
Q'_j}]}\it{A}_w({\bf{r,Q,K}})\it{B}_w({\bf{r,Q',K'}})
\end{eqnarray}

Quite similarly we obtain that $\it{BA}$ is
\begin{eqnarray}\label{31}
\it{B
A}=d{\bf{Q}}d{\bf{K}}d{\bf{Q'}}d{\bf{K'}}\it{A}_w({\bf{r,Q,K}})\it{B}_w({\bf{r,Q',K'}})
F({\bf{r,Q,K}})F({\bf{r,Q',K'}})=\nonumber \\
\int
d{\bf{Q}}d{\bf{K}}d{\bf{Q'}}d{\bf{K'}}\it{A}_w({\bf{r,Q,K}})\it{B}_w({\bf{r,Q',K'}})e^{(i/4\pi)
\sum[\frac{\partial}{\partial Q_j}\frac{\partial}{\partial
K'_j}-\frac{\partial}{\partial K_j}\frac{\partial}{\partial
Q'_j}]}\times\nonumber \\ e^{(i/4\pi)\sum(1+2 r_j)
[\frac{\partial}{\partial Q_j}\frac{\partial}{\partial
K'_j}+\frac{\partial}{\partial K_j}\frac{\partial}{\partial
Q'_j}]}F({\bf{r,Q,K}})\delta({\bf{Q'-Q}})\delta({\bf{K'-K}})=\nonumber
\\ \int
d{\bf{Q}}d{\bf{K}}d{\bf{Q'}}d{\bf{K'}}F({\bf{r,Q,K}})\delta({\bf{Q'-Q}})
\delta({\bf{K'-K}})e^{-(i/4\pi) \sum[\frac{\partial}{\partial
Q_j}\frac{\partial}{\partial K'_j}-\frac{\partial}{\partial
K_j}\frac{\partial}{\partial Q'_j}]}\times\nonumber \\
e^{(i/4\pi)\sum(1+2 r_j) [\frac{\partial}{\partial
Q_j}\frac{\partial}{\partial K'_j}+\frac{\partial}{\partial
K_j}\frac{\partial}{\partial
Q'_j}]}\it{A}_w({\bf{r,Q,K}})\it{B}_w({\bf{r,Q',K'}})
\end{eqnarray}
Therefore the commutator $[\it{A,B}]$ is
\begin{eqnarray}\label{32}
{[\it{A,B}]=2i\int
d{\bf{Q}}d{\bf{K}}d{\bf{Q'}}d{\bf{K'}}F({\bf{r,Q,K}})\delta({\bf{Q'-Q}})
\delta({\bf{K'-K}})}\nonumber \\e^{(i/4\pi)\sum(1+2 r_j)
[\frac{\partial}{\partial Q_j}\frac{\partial}{\partial
K'_j}+\frac{\partial}{\partial K_j}\frac{\partial}{\partial
Q'_j}]}sin\{\frac{1}{4\pi}[\frac{\partial^2}{\partial{\bf{Q}}\partial{\bf{K'}}}-\frac{\partial^2}
{\partial{\bf{Q'}}\partial{\bf{K}}}]\}\it{A}_w({\bf{r,Q,K}})\it{B}_w({\bf{r,Q',K'}})
\end{eqnarray}

The quantum equation of motion for an operator $\rho$ is
\begin{equation}\label{33}
-i\hbar\frac{\partial \it{ \rho}}{\partial t}=[\it{\rho},H]
\end{equation}
where the operator $\rho$ is given by Eq.(12)
\begin{equation}\label{34}
\it{\rho}=\int
d{\bf{Q}}d{\bf{K}}\it{\rho}_w(t,{\bf{r,Q,K}})F({\bf{r,Q,K}})
\end{equation}
Inserting (32) and (34) into (33) we obtain
\begin{eqnarray}\label{35}
\frac{\partial \it{ \rho}_w({\bf{r,Q,P}})}{\partial t}=\nonumber
\\
\frac{2}{\hbar}\int
d{\bf{Q'}}d{\bf{K'}}\delta({\bf{Q'-Q}})\delta({\bf{K'-K}})
e^{\frac{i}{4\pi}\sum(1+2r_j)[\frac{\partial^2}{\partial
Q'_j\partial K_j}+\frac{\partial^2}{\partial Q_j\partial
K'_j}]}\times \nonumber \\
sin\{\frac{1}{4\pi}[\frac{\partial^2}{\partial Q'_j\partial
K_j}-\frac{\partial^2}{\partial Q_j\partial
K'_j}]\}\it{H}_w({\bf{Q'}},{\bf{K'}})\it {\rho}_w(t,{\bf{Q,K}})
\end{eqnarray}
Performing integration  and replacing parameter ${\bf{r}}$ by the
following $$r_j=-\frac{(1+s_j)}{2},~~ j=1,2,...,N$$ we obtain the
generalization of the Moyal bracket \cite{JM}:
\begin{equation}\label{36}
\frac{\partial \it{ \rho}({\bf{r,Q,P}})}{\partial
t}=[\it{H},\it{\rho}]_M^s
\end{equation}
where ( after returning to the variables  $P_j=2\pi\hbar K_j$) the
generalized (or $s$-parameterized) Moyal bracket is
$$[\it{H},\it{\rho}]_M^s=e^{-\frac{i\hbar}{2}\sum s_j(\frac{\partial^2}
{\partial Q'_j\partial P_j}+\frac{\partial^2}{\partial Q_j\partial
P'_j})}\times $$
$$\frac{2}{\hbar}sin\{\frac{\hbar}{2}[\frac{\partial^2}{\partial Q'_j\partial
P_j}-\frac{\partial^2}{\partial Q_j\partial
P'_j}]\}\it{H}_w({\bf{Q'}},{\bf{P'}})\it
{\rho}_w(t,{\bf{Q,P}})$$\\

The same result was presented in \cite{DW}. However there the
authors introduced parameter $s$ by hand, without relating it to
any transformation of the quantum states and simply treating it as
a means to achieve a unified approach to different quantization
rules. On the other hand, our approach ( based on \cite{BL})
explicitly shows that such  a parameter is a result of a linear
transformation from one quantum state to another. Since the pure
states are represented by rays, it is very natural to expect that
the above transformation would result in the appearance of the
phase, which is clearly seen in the exponential
operator.\\
\section{On a Physical Meaning\\
 of the s-Parameter}
It is therefore interesting to investigate what role is played by
this parameter in a transition to a classical case. To this end we
restrict our attention to a $1-D$ case and consider (following
Moyal) space-conditional moments $<p^n>_w$ ( the Wigner averages
of the powers of $p^n$ of the momentum):
\begin{equation}\label{37}
<p^n>_w=\frac{\int A_w(p,q,s)p^ndp}{\int A_w(p,q,s)dp}
\end{equation}

For the subsequent calculations we have to transform $A_w$ (the
phase-space distribution function given by Eq.\ref{28}) into an
integral form. For the convenience sake, we replace parameter $s$
by $-\sigma$ and choose the units with $\hbar=1$. As a next step,
we find Fourier-transform of $A_w(p,q,\sigma)$, Eq.(\ref{28}). We
denote this transform by $ M(\tau,\theta,\sigma)$ :
\begin{equation}
\label{38} M(\tau,\theta,\sigma)=\int\int e^{(\tau p+\theta
q)}e^{-\frac{1-\sigma}{2}\frac{\partial^2}{\partial p
\partial q}}[\Psi^*(q)\Psi(p)e^{ipq}]dpdq
\end{equation}
Integration of (\ref{38}) by parts yields:
\begin{equation}
\label{39} M(\tau,\theta,\sigma)=\int\int\Psi^*(q)~
e^{i\theta[q+\frac{\tau (1-\sigma)}{2}]}~\Psi(p)e^{ip(q+
\tau)}dpdq
\end{equation}\\

We introduce a new variable $q_1$:
$$q=q_1-\frac{\tau}{2}(1-\sigma)$$
Using $q_1$ in (\ref{39}) we obtain:
\begin{equation}
\label{40} M(\tau,\theta,\sigma)=\int\{ e^{i\theta q_1}
\Psi^*[q_1-\frac{\tau}{2}(1-\sigma)] ~\int \
e^{ip[q_1+\frac{\tau}{2}(1+\sigma)]}~\Psi(p)dp\}dq_1
\end{equation}
Since
$$\frac{1}{\sqrt\hbar}\int
e^{ip[q_1+\frac{\tau}{2}(1+\sigma)]}\Psi(p)dp=
\Psi[q_1+\frac{\tau}{2}(1+\sigma)]$$

relation (\ref{40}) takes the following form:
\begin{equation}
\label{41}
M(\tau,\theta,\sigma)=\int\Psi^*[q_1-\frac{\tau}{2}(1-\sigma)]
e^{i\theta q_1}\Psi[q_1+\frac{\tau}{2}(1+\sigma)]dq_1
\end{equation}
Inverse Fourier-transform of $M(\tau,\theta,\sigma)$ gives us the
desired integral form of the phase-space distribution (a
$s$-parameterized Wigner function) $A_w(p,q,\sigma)$\footnote{ a
simplified derivation of $A_w(p,q,\sigma$ is given in the
Appendix}:
\begin{equation}
\label{42}A_w(p,q,\sigma)
=\frac{1}{2\pi}\int\Psi^*[q_1-\frac{\tau}{2}(1-\sigma)]~ e^{-i\tau
p}~\Psi[q_1+\frac{\tau}{2}(1+\sigma)]d\tau
\end{equation}
If parameter $\sigma =0$ then , as expected, Eq.(\ref{42}) is
reduced to an expression first given by Wigner \cite{EW}. Using
Eq. (\ref{42}) we find the probability density
\begin{equation}
\label{43} \int
A_w(p,q,\sigma)dp=\Psi^*(q)\Psi(q)=|\Psi(q)|^2\equiv\rho
\end{equation}\\

In general, since the parameter $\sigma$ (or $s$) is
complex-valued, the $\sigma$-parameterized Wigner function
$A_w(p,q,\sigma)$ is also $complex-valued$, in contradistinction
to its conventional counterpart ( with $\sigma=0$). However for
the purely imaginary values of the parameter $\sigma$, the
$\sigma$-parameterized Wigner function becomes real-valued again:
$$i)~~~~A_w^*(p,q,\sigma)=A_w(q,p,-\sigma)$$
$$ii)~~~~A_w^*(p,q,\sigma)=A_w(q,p,\sigma),~~~~ Re(\sigma)=0$$\\

 For the following we rewrite $A_w(p,q,\sigma)$ in terms of the
momentum wave function $\Phi(p)$. After some algebra we obtain
\begin{equation}
\label{44} A_w(p,q,\sigma)=\int\int
dp'dp''e^{-iq(p''-p')}\Phi^*(p'')\Phi(p')\delta[p-\frac{p''(1+\sigma)+p'(1-\sigma)}{2}]
\end{equation}
Upon substitution of (\ref{43}), (\ref{44}) into (\ref{37}) we get
\begin{eqnarray}
\label{45} \rho<p^n>_w=\int A_w(p,q,\sigma) p^n dp=   \nonumber \\
\int\int\int dp dp'dp''p^n e^{-iq(p''-p')}\Phi^*(p'')
\Phi(p')\delta[p-\frac{p''(1+\sigma)+p'(1-\sigma)}{2}]= \nonumber\\
\int\int dp'dp'' e^{-iq(p''-p')}\Phi^*(p'')
\Phi(p')[\frac{p''(1+\sigma)+p'(1-\sigma)}{2}]^n
\end{eqnarray}
By observing that
$$\frac{1-\sigma}{i}\frac{\partial}{\partial q_2}e^{-i(q_1p''-q_2p')}=
(1-\sigma)p'e^{-i(q_1p''-q_2p')}|_{q_1\rightarrow
q_2}=(1-\sigma)e^{-iq(p'-p'')}$$
$$-\frac{1+\sigma}{i}\frac{\partial}{\partial q_1}e^{-i(q_1p''-q_2p')}=
(1-\sigma)p''e^{-i(q_1p''-q_2p')}|_{q_1\rightarrow
q_2}=(1+\sigma)e^{-iq(p'-p'')}$$ we rewrite (\ref{45})
\begin{eqnarray}
\label{46} \rho
<p^n>_w=\{\frac{1}{2i}[(1-\sigma)\frac{\partial}{\partial
q_2}-(1+\sigma)\frac{\partial}{\partial q_1}]^n\int
e^{-iq_1p'}\Phi^*(p'')dp''\int e^{iq_2p'}\}|_{q_2 \rightarrow
q_1}=\nonumber\\
\{\frac{1}{2i}[(1-\sigma)\frac{\partial}{\partial
q_2}-(1+\sigma)\frac{\partial}{\partial
q_1}]^n\Psi^*(q_1)\Psi(q)\}|_{q_2 \rightarrow q_1}
\end{eqnarray}

Returning to the units with $\hbar$ and using (\ref{43}), we
calculate two first momenta $<p>_w$ and $<p^2>_w$:
\begin{eqnarray}
\label{47} <p>_w=\frac{1}{\Psi^*(q)\Psi(q)}\{\frac{\hbar}{2i}
[(1-\sigma)\frac{\partial}{\partial
q_2}-(1+\sigma)\frac{\partial}{\partial
q_1}]\Psi^*(q_1)\Psi(q)\}|_{q_2 \rightarrow q_1}=\nonumber \\
\frac{\hbar}{2i}\frac{\partial}{\partial
q}\{Ln(\frac{\Psi}{\Psi^*})-\sigma Ln(\Psi^*\Psi)\}
\end{eqnarray}
and
\begin{equation}
\label{48}
<p^2>_w=-\frac{\hbar^2}{4}[(1-\sigma)^2\frac{\Psi''}{\Psi}-2(1-\sigma^2)
\frac{\partial Ln\Psi}{\partial q}\frac{\partial
Ln\Psi^*}{\partial q}+(1+\sigma)^2\frac{\Psi^{*''}}{\Psi^*}]
\end{equation}\\

Let us consider a semi-classical limit
\begin{equation}
\label{49} \Psi(q,t)=\sqrt{\rho}e^{iS/\hbar}
\end{equation}
where $S$ is the classical action. Inserting (\ref{49}) into
(\ref{47})  we obtain:
\begin{equation}
\label{50} \lim_{\hbar \rightarrow 0}<p>_w = \lim_{\hbar
\rightarrow 0}\{\nabla S-i\frac{\hbar S}{2} \nabla(Ln
\rho)\}=\nabla S=p_{classical}
\end{equation}
If we use the Schroedinger equation:
$$i\hbar\frac{\partial Ln(\Psi)}{\partial t}=-\frac{\hbar^2}{2m}\frac{\nabla^2 \Psi}{\Psi}+V$$
then employing (\ref{49}) in (\ref{48})we get the following:
\begin{eqnarray}
\label{51} \lim_{\hbar \rightarrow 0}<p^2>_w=m\lim_{\hbar
\rightarrow 0}\{-\frac{\partial S}{\partial
t}-V+\frac{1}{2m}(\nabla S)^2+\frac{\hbar^2}{8m}(\nabla
Ln\rho)^2+\nonumber \\
+\sigma^2[-\frac{\partial S}{\partial t}-V-\frac{1}{2m}(\nabla
S)^2-\frac{\hbar^2}{8m}(\nabla
Ln\rho)^2]+\frac{i\hbar}{2}\frac{\partial}{\partial
t}Ln(\frac{\Psi}{\Psi^*})\}=\nonumber\\
m\{-\frac{\partial S}{\partial t}-V+\frac{1}{2m}(\nabla
S)^2+\frac{\hbar^2}{8m}(\nabla Ln\rho)^2]+\nonumber \\
\sigma^2[-\frac{\partial S}{\partial t}-V-\frac{1}{2m}(\nabla
S)^2-\frac{\hbar^2}{8m}(\nabla Ln\rho)^2\}
\end{eqnarray}
This limit must yield the classical value of the square of the
classical momentum
$$\lim_{\hbar \rightarrow 0}<p^2>_w=p^2_{classical}=(\nabla S)^2$$
which is independent of the parameter $\sigma.$ This is possible
if the factor at $\sigma$ in (\ref{51}) becomes $0$, that is
$$-\frac{\partial
S}{\partial t}=\frac{1}{2m}(\nabla S)^2+V$$ But amazingly enough
this condition is nothing more than the classical Hamilton-Jacobi
equation. Thus emergence of the parameter $\sigma$ in Wigner
function is tied to an emergence of the classical Hamilton-Jacobi
equation in transition to a classical regime.
\section{Conclusion}
We have demonstrated that $s$-parameterized Wigner function
emerges as a result of a linear transformation from one quantum
state to another. The respective change of the phase space
coordinates accompanying such a transform must be necessarily
canonical. This allows one to arrive in a natural way, without
introducing "by  hand" the $s$-parameter into the transformation
either by suitably chosen displacement operator or by using it as
a "missing link" between normal and anti-normal ordering
operators.

A transition to a classical regime demonstrates that parameter $s$
plays an important role, ensuring the emergence of the classical
Hamilton-Jacobi equation as a condition for the disappearance of
this parameter in classical mechanics.
\section{Appendix}
Since the momentum representation $\Phi(p)$ and the coordinate
representation $\Psi(q)$ of the wave function ( for simplicity
sake we consider a 1-D case) are related  as follows:
\begin{equation}
\label{a1} \Phi(p)=\frac{1}{\sqrt{2\pi\hbar}}\int
e^{-ipq'/\hbar}\Psi(q')dq'
\end{equation}
the momentum probability density is
\begin{equation}
\label{a2} |\Psi(p)|^2=\frac{1}{2\pi\hbar}\int\int dq'dq''
\Psi(q')\Psi^*(q'')e^{ip(q''-q')/\hbar}
\end{equation}
We introduce new variables $q$ and $\tau$
$$q'=q+\alpha\tau;~~~q''=q+\beta\tau$$
where $\alpha$ and $\beta$ are constants such that the Jacobian of
transformation from $q'',q'$ to $q,\tau$ is $\hbar$ which means
$$\alpha-\beta =\hbar$$ As  a result, Eq.(\ref{a2}) yields
\begin{equation}
\label{a3} |\Phi(p)|^2 =\frac{1}{2\pi}\int\int dq
d\tau\Psi(q+\alpha\tau)\Psi^*[q+(\alpha-\hbar)\tau]e^{-ip\tau}
\end{equation}\\

Now we represent the arbitrary parameter $\alpha$ as
$$\alpha=\frac{\hbar(1+s)}{2}$$
Upon substitution of this expression into (\ref{a3}) we obtain
\begin{eqnarray}
\label{a4} |\Phi(p)|^2=\frac{1}{2\pi}\int dq\int d\tau
e^{-ip\tau}\Psi[q+\frac{\hbar(1+s)}{2}]\Psi^*[q-\frac{\hbar(1-s)}{2}]=
\nonumber\\
\int dq A_w(p,q,s)
\end{eqnarray}
where
\begin{equation}
\label{a5} A_w(p,q,s)=\frac{1}{2\pi}\int
\Psi[q+\frac{\hbar(1+s)}{2}]e^{-ip\tau}\Psi^*[q-\frac{\hbar(1-s)}{2}]
d\tau
\end{equation}
is the $s$-parameterized Wigner function found earlier,
Eq.(\ref{42})\\

\section{Acknowledgement}
The author expresses his gratitude to C.McCallum for his help in
preparation of this paper.

\end{document}